\documentclass[10pt,journal]{IEEEtran}
\usepackage[utf8]{inputenc}
\usepackage[T1]{fontenc}
\usepackage{microtype}
\usepackage{cite}
\usepackage[caption=false]{subfig}
\usepackage{amsmath, amsfonts, amsthm, amssymb}
\usepackage{algorithm}
\usepackage{algpseudocode}
\usepackage{graphicx}
\usepackage{tikz}
\usetikzlibrary{automata, positioning, decorations.pathreplacing, shapes.geometric}
\usepackage{mathtools}
\usepackage{commath}
\usepackage{mathptmx}
\usepackage{hyperref}
\hypersetup{
    colorlinks = true,
    linkcolor  = blue,
    citecolor  = red,
    filecolor  = magenta,
    urlcolor   = blue,
}

\theoremstyle{definition}

\theoremstyle{remark}

\theoremstyle{plain}

\pdfoutput=1

\begin{document}

\title{Adapting Large Language Models for Improving TCP Fairness over WiFi}

\author{
    Shyam Kumar Shrestha, Shiva Raj Pokhrel, and Jonathan Kua%
    \IEEEcompsocitemizethanks{
        \IEEEcompsocthanksitem S.K. Shrestha, S.R. Pokhrel, and J. Kua are with the School of Information Technology, Deakin University, Geelong, Australia.
        Email: \{shyam.shrestha, shiva.pokhrel, jonathan.kua\}@deakin.edu.au
    }%
}

\maketitle

\begin{abstract}
The new transmission control protocol (TCP) relies on Deep Learning (DL) for prediction and optimization, but requires significant manual effort to design deep neural networks (DNNs) and struggles with generalization in dynamic environments. Inspired by the success of large language models (LLMs), this study proposes TCP-LLM, a novel framework leveraging LLMs for TCP applications. TCP-LLM utilizes pre-trained knowledge to reduce engineering effort, enhance generalization, and deliver superior performance across diverse TCP tasks. Applied to reducing flow unfairness, adapting congestion control, and preventing starvation, TCP-LLM demonstrates significant improvements over TCP with minimal fine-tuning.
\end{abstract}


\section{Introduction} 
\label{sec:introduction}

Since its invention, TCP optimization has been dominated by rule-based algorithms built on mathematical formulations. Traditional algorithms such as Reno and Cubic~\cite{ha2008cubic} rely on predefined models to adjust congestion windows based on metrics like packet loss and end-to-end delays. Recent advancements, including BBR~\cite{cardwell2017bbr} and PCC~\cite{dong2018pcc}, have bandwidth estimation and utility-based models. 

Despite these advancements, conventional approaches face significant challenges in adapting to modern, heterogeneous networks. Today’s networks, encompassing technologies like WiFi, 5G/6G, and satellite systems, exhibit high variability and dynamic conditions that often exceed the adaptive capacity of traditional TCP algorithms. Moreover, these formulaic methods require extensive parameter tuning, making them computationally expensive and less effective in diverse and evolving networks~\cite{Pokhrelharness2021, shrestha2024fairness}.

Machine learning (ML), particularly Deep Learning (DL) and Deep Reinforcement Learning (DRL), has emerged as a transformative solution for TCP optimization. Unlike traditional methods, ML-driven approaches dynamically adapt to complex network conditions without relying heavily on manual parameter tuning. Supervised Learning (SL) has been successful in tasks like traffic classification and bandwidth prediction~\cite{mei2020realtime}, while DRL excels in decision-making tasks such as congestion window optimization, adaptive load balancing, and real-time Congestion Control Algorithm (CCA) selection. DRL has demonstrated its ability to address critical issues in TCP systems, such as unfairness and inefficiencies, and has extended its impact to broader networking applications, including adaptive bitrate streaming~\cite{kan2022improving} and cloud resource allocation~\cite{mao2019learning}. Our prior work~\cite{shrestha2024fairness} highlights the potential of DRL in dynamically optimizing TCP performance, ensuring fairness and mitigating starvation under varying network conditions.

The emergence of Large Language Models (LLMs), such as GPT-4~\cite{achiam2023gpt} and Llama2~\cite{touvron2023llama2}, introduces new possibilities for TCP optimization. LLMs exhibit remarkable capabilities in reasoning, adaptability, and generalization, which extend beyond traditional natural language processing tasks to domains like robotics, protein design, and decision-making in complex systems~\cite{zhang2024mm, lin2023evolutionary}. These capabilities make LLMs a compelling candidate for addressing TCP challenges in heterogeneous and dynamic networks. By dynamically interpreting network metrics and adapting CCAs, LLMs have the potential to simplify engineering complexities, minimize manual tuning, and optimize performance in real time.

However, integrating LLMs into TCP presents unique challenges. Cross-modality mismatches, delayed decision processes, and resource-intensive adaptation are significant barriers to their effective application in this domain~\cite{zhang2022opt, li2023halueval, lin2023evolutionary}. To address these challenges, we propose TCP-LLM, a novel framework that combines the strengths of LLMs with efficient adaptation techniques to revolutionize TCP optimization. This paper explores how TCP-LLM overcomes these limitations, providing a robust solution to improve network performance in the diverse and evolving environments of today. TCP-LLM is a pioneering framework that bridges the adaptability of LLM with the performance requirements of TCP.\\ Our key contributions are as follows.\\
 \textit{1. A comprehensive analysis of the challenges in adapting LLMs to TCP optimization, focusing on how network-specific metrics like RTT, throughput, and loss rate impact LLM performance and the complexities of real-time integration.\\
2.  Development of the TCP-LLM framework, which seamlessly integrates LLM into TCP to improve fairness, prevent starvation, and dynamically select CCA.\\
3. Introduction of efficient, parameter-optimized techniques for tailoring LLMs to TCP-specific tasks, reducing fine-tuning costs while maintaining adaptability. 4.  Rigorous experimentation showcasing the efficacy of TCP-LLM in diverse and dynamic network environments, demonstrating significant improvements in performance metrics.}

\section{Background and Related Work}\label{Literature Review}
SL has emerged as a valuable tool for predictive tasks in networks, including traffic classification, bandwidth forecasting~\cite{mei2020realtime}, and viewport prediction~\cite{rondon2021track}. In TCP optimization, SL can be applied to predict congestion patterns and traffic conditions, enabling proactive adjustments to enhance performance. For instance, Yan et al.~\cite{yan2020learning} demonstrated the potential of deep neural networks (DNNs) by predicting future bandwidth in video streaming scenarios, thereby improving bitrate quality through anticipatory decision-making.

Reinforcement learning, particularly DRL, is well-suited for sequential decision-making tasks, such as congestion control~\cite{yen2023computers}, adaptive streaming, and resource allocation. Unlike SL, DRL interacts with its environment to learn optimal strategies through iterative feedback and rewards. This capability has led to the development of adaptive TCP algorithms that adjust to dynamic network conditions. For example, Mao et al.~\cite{mao2019learning} applied RL to optimize resource allocation in distributed computing, while Shrestha et al.~\cite{shrestha2024fairness} employed DRL to enhance TCP fairness by dynamically switching CCA.

Despite the potential of ML, challenges remain. The development of DRL models is highly labor-intensive, requiring significant human effort to design architectures, tune reward functions, and optimize learning processes~\cite{meng2020interpreting}. Furthermore, these models often lack generalization, performing poorly in unseen or highly heterogeneous network conditions, such as WiFi or 5G/6G environments~\cite{shrestha2024fairness}. High computational overhead further limits scalability, as model training and fine-tuning demand extensive resources~\cite{wu2024netllm}. These limitations hinder the widespread adoption of ML-based TCP solutions in real-world scenarios.

LLM, such as GPT-4~\cite{achiam2023gpt} and Llama2~\cite{touvron2023llama2}, offer new possibilities for overcoming these challenges. LLMs, built on Transformer architectures~\cite{vaswani2017attention}, have demonstrated exceptional capabilities in understanding and generating complex sequences, extending their applications far beyond natural language processing. Pre-trained on massive datasets, LLMs exhibit remarkable adaptability, reasoning, and contextual understanding, enabling use cases ranging from dialogue generation to task-specific fine-tuning~\cite{li2024dollm}.

The structure of LLMs relies on tokenized representations of input sequences, which are processed through multiple layers of attention-based mechanisms to capture dependencies and generate predictions. This architecture allows LLMs to generalize across diverse tasks, making them highly effective in domains like robotics~\cite{zhang2024mm}, healthcare~\cite{lin2023evolutionary}, and network traffic analysis~\cite{wu2024netllm}. However, adapting LLMs for TCP introduces unique challenges, including the need to process numerical inputs such as RTT and loss rates, handle real-time constraints, and mitigate resource-intensive fine-tuning requirements~\cite{zhang2022opt, li2023halueval}.

Motivated by these capabilities, this paper explores the adaptation of LLMs to TCP optimization through the proposed TCP-LLM framework. While prior research has highlighted the potential of LLMs in related fields~\cite{li2023chattwin}, systematic frameworks tailored to address TCP challenges remain underdeveloped. TCP-LLM bridges this gap by integrating LLMs with efficient adaptation techniques, enabling dynamic CCA selection, fairness enhancement, and real-time optimization in diverse and evolving network environments.

\section{Research Motivation and Relevance} \label{motivation}

This section focuses on three critical TCP tasks—flow fairness, starvation prevention, and CCA compatibility—that are central to TCP-LLM implementation. Flow fairness ensures equitable bandwidth distribution by monitoring metrics like throughput, packet loss, and RTT~\cite{arun2022starvation}, preventing resource monopolization~\cite{shrestha2024fairness}. Starvation prevention addresses disproportionate bandwidth allocation to maintain fairness and enhance reliability~\cite{philip2021revisiting}. CCA compatibility ensures coexistence among algorithms like Cubic, BBR, and PCC, preventing resource domination and flow unfairness~\cite{yamazaki2021fairness}.

These tasks integrate SL for historical data analysis and reinforcement learning (RL) for real-time decisions~\cite{afifi2024machine}. They reflect distributed TCP control, where connections autonomously optimize performance, and leverage scalar and time-series data to address dynamic challenges~\cite{yen2023computers}. Their relevance and complexity make them ideal for demonstrating the capabilities of TCP-LLM.

\begin{table*}[t]
\centering
\small 
\renewcommand{\arraystretch}{1.1} 
\begin{tabular}{|p{2.2cm}|p{3.8cm}|p{3.8cm}|p{3.8cm}|}
\hline
\textbf{Features} & \textbf{Flow Fairness} & \textbf{Starvation Prevention} & \textbf{CCA Compatibility} \\ \hline
\textbf{Input} & Throughput, loss rate, RTT & RTT \& throughput trends, loss rate & Loss rate, RTT, throughput \\ \hline
\textbf{Output} & Selected CCA, optimized TCP parameters & Balanced TCP metric trends & Predicted compatible CCA \\ \hline
\textbf{Learning Paradigm} & RL & RL & SL, RL \\ \hline
\textbf{Data Modality} & Scalar & Time-series & Scalar \& Time-series \\ \hline
\textbf{Use Case} & Ensure equitable bandwidth sharing & Address flow starvation issues & Identify and integrate compatible CCA \\ \hline
\end{tabular}
\vspace{0.2cm} 
\caption{Key Features and Use Cases for TCP-LLM in Addressing TCP Tasks}
\label{table1}
\end{table*}

Table~\ref{table1} summarizes the key TCP-LLM tasks, detailing their features, including input data, outputs, learning paradigms, data modalities, and use cases. It highlights the approaches for tasks like flow fairness, starvation prevention, and CCA compatibility, emphasizing how TCP-LLM utilizes scalar and time-series data to address diverse networking challenges effectively. This breakdown illustrates the practical goals achieved in optimizing network performance through SL and RL. However, adapting language models to TCP presents challenges.

A significant challenge is the \textit{cross-modality mismatch}, where TCP tasks rely on numerical and time-series data such as throughput, loss rate, and RTT, while LLMs are designed for text processing. This disparity, termed the \textit{modality gap}, complicates direct application of LLMs to TCP tasks, requiring innovative methods to bridge this divide effectively.

     \begin{figure}[h]
     \centering
     \subfloat[Answer Validiy]{
     \includegraphics[scale=0.3757]{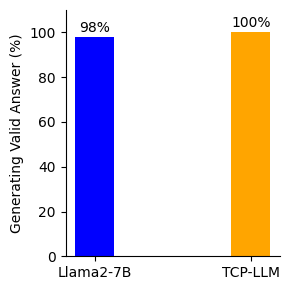}
     \label{fig:answer_val}
     }
   \hfill
    \subfloat[Answer Generation]{
    \includegraphics[scale=0.3757]{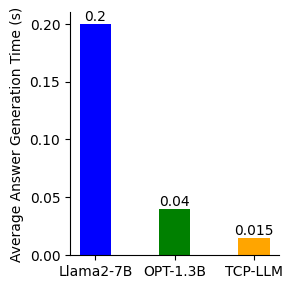}
    \label{fig:answer_gen}
     }
   \caption{Visualization of performance efficiency and reliability of LLMs in comparison to TCP-LLM in CCA selection TCP task}
   \label{fig:Answer}
   \end{figure}

One approach to addressing the \textit{modality gap} is \textbf{prompt learning}~\cite{liu2023cav3}, which converts numerical data into textual formats using predefined templates to enable LLMs to interpret and respond effectively. For instance, TCP metrics like RTT and throughput can be represented as textual descriptions (e.g., \textit{throughput is 100 Mbps; RTT is 50 ms over the past second}). While this approach has succeeded in other domains, it faces limitations in TCP tasks.

The dynamic and temporal nature of TCP data, such as fluctuating RTT or throughput trends, often cannot be fully captured through textual transformations. For example, when Llama2-7B was adapted using prompt-based methods to predict TCP metrics from historical data, its performance was suboptimal (see Figure~\ref{fig:Answer}) compared to specialized approaches for numerical and time-series data. This shortfall arises because text-based formats oversimplify the intricate temporal dependencies critical for accurate predictions. Consequently, the model struggles to achieve precision, underscoring the modality gap as a significant barrier to effectively applying LLMs in TCP tasks.

The auto-regressive nature of LLMs' response generation, illustrated in presents significant challenges for real-time TCP applications. One major issue is the uncertainty in token prediction, often leading to physically invalid outputs, a phenomenon known as hallucination~\cite{ji2023survey}. For instance, evaluations of Llama2's predictions for TCP tasks revealed instances of invalid outputs (Figure~\ref{fig:answer_val}), undermining reliability in scenarios where precision is critical, \textit{delayed decision process}. To address this, TCP-LLM introduces mechanisms that ensure prediction validity, bridging the gap left by traditional LLMs. Another challenge is the latency associated with tokenization. LLMs decompose responses into multiple tokens, requiring several inference steps to complete a single response. As shown in Figure~\ref{fig:answer_gen}, Llama2 takes an average of 0.2 seconds to produce responses, which exceeds the minimal response times required for real-time decision-making~\cite{chen2018auto}. This delay compromises efficiency and reliability in applications demanding rapid adaptation, highlighting a critical limitation for using LLMs in real-time TCP scenarios.


     \begin{figure}[h]
     \centering
     
    \includegraphics[width=3.4025 in]{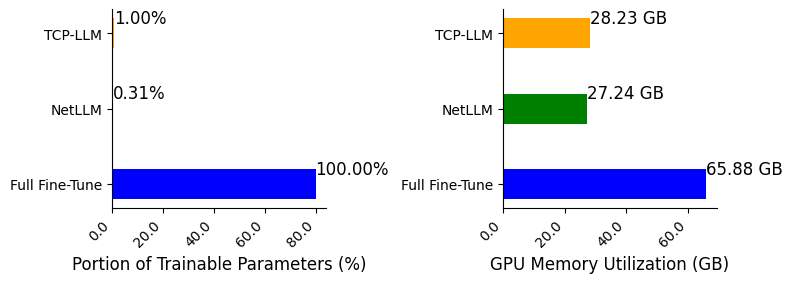}
    
    \caption{Comparision of the trainable parameters and resource-intensive adaptation cost of a whole parameter fine-tuning. TCP-LLM's low-rank TCP adaptation approach minimizes the cost of implementing LLM in TCP.}
   \label{fig:cost minimization}
   \end{figure}

TCP-related tasks like flow fairness, starvation prevention, and CCA compatibility often rely on RL to optimize complex system behaviors. These tasks require active interaction with dynamic network environments, characterized by fluctuating bandwidth and workloads, to collect data and optimize rewards. RL-based methods, such as Proximal Policy ~\cite{zheng2023secrets, wu2023pairwise}, commonly used to fine-tune LLM, are inherently resource intensive. This limitation becomes particularly evident, \textit{resource-intensive adaptation}, when adapting models like Llama2 to TCP tasks.

Fine-tuning Llama2 for TCP requires numerous iterations involving extensive data collection and reward optimization, which accounts for over 50\% of the total training time~\cite{wu2024netllm}. The computational burden is further amplified when full-parameter fine-tuning is applied. For example, we observe that adapting Llama2-7B for CCA compatibility demands 65.88 GB of GPU memory with 100\% of trainable parameters, as shown in Figure~\ref{fig:cost minimization}. This high resource consumption is due to the model's large parameter size, requiring substantial memory and computational resources, which is evident by the 1\% parameters with a GPU of 28.23 GB.

\section{Our TCP-LLM Design}\label{model design}

This section introduces the comprehensive design of TCP-LLM, a framework tailored to adapt LLMs for addressing the TCP-related challenges discussed earlier. The framework comprises three primary components: the Integrated Encoder, the TCP-LLM Head, and Low-Rank TCP Adaptation. These components collectively enable efficient processing of TCP-specific data and enhance model adaptability.

\begin{figure} [!h]
    \centering
    \includegraphics[scale=0.344]{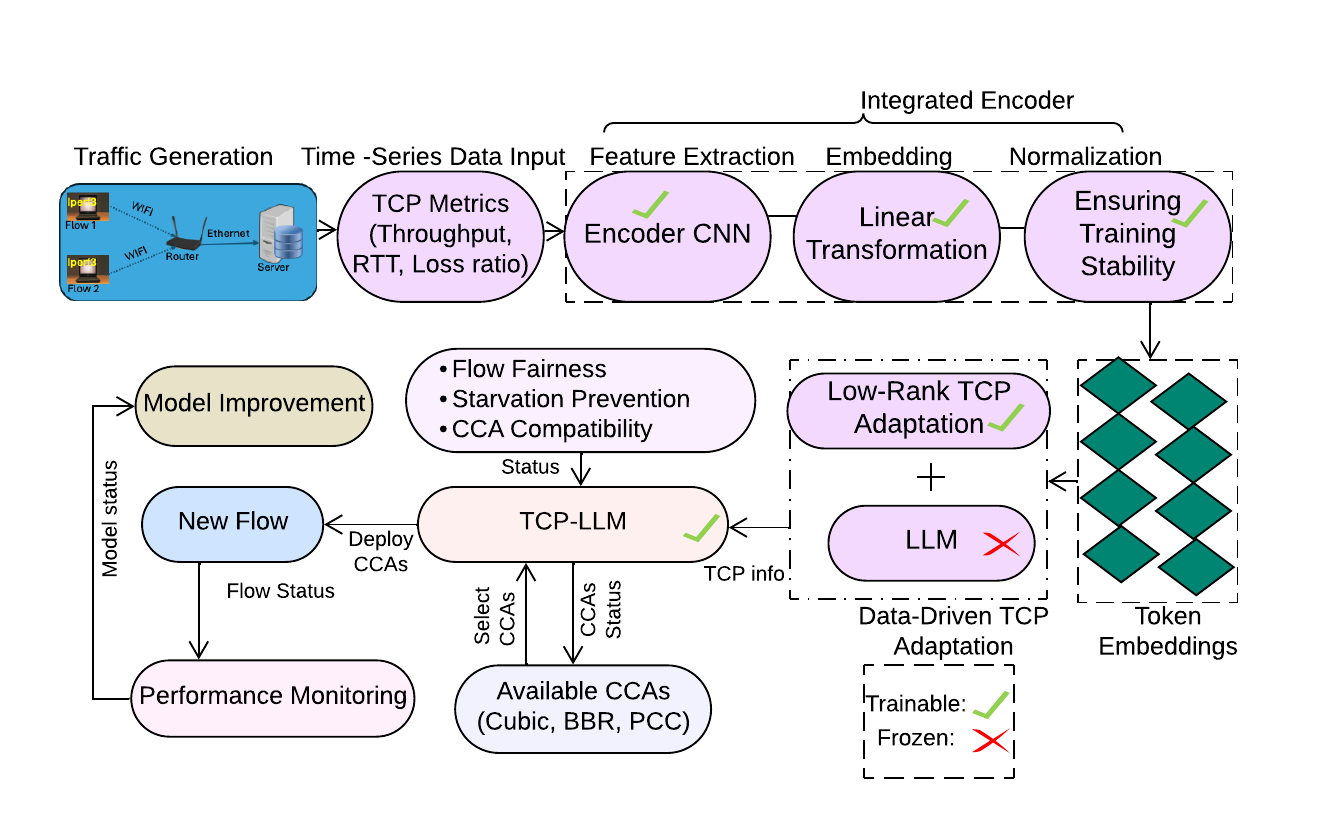}
    \caption{TCP-LLM features an encoder for data processing, a task-specific head, and Low-Rank Adaptation for efficient learning, with extensible principles demonstrated on three key TCP tasks.}
    \label{fig:tcp-llm}
\end{figure}

Before detailing these core elements, we present an overview of the framework's design and implementation. Figure~\ref{fig:tcp-llm} illustrates the architecture of TCP-LLM, while Algorithm~\ref{design} outlines the step-by-step implementation of the framework for TCP tasks. This combination of visual and algorithmic representation provides a clear and comprehensive understanding of TCP-LLM's functionality and its application to flow fairness, starvation prevention, and CCA compatibility.

\begin{algorithm} 
\caption{TCP-LLM Operation}\label{design}
\begin{algorithmic}[1]
\State \textbf{Initialize} TCP-LLM

\State \textbf{Step 1:} Collect historical metrics
\State \texttt{hist\_metrics} = \texttt{collect\_metrics(source)}

\State \textbf{Step 2:} Encode time-series data with CNN
\State \texttt{enc\_features} = \texttt{encode\_CNN(hist\_metrics)}

\State \textbf{Step 3:} Linear transform and normalize features
\State \texttt{tok\_emb} = \texttt{lin\_transform(enc\_features)}
\State \texttt{norm\_emb} = \texttt{normalize(tok\_emb)}

\State \textbf{Step 4:} Analyze flow fairness
\State \texttt{flow\_stat} = \texttt{analyze\_fairness(norm\_emb)}

\State \textbf{Step 5:} Assess fairness
\If {\texttt{detect\_unfairness(flow\_stat)}}
    \State \textbf{Step 6:} Adjust for fairness
    \State TCP-LLM.adjust\_CCA\_fairness(flow\_stat)
\EndIf

\State \textbf{Step 7:} Check for starvation
\If {\texttt{detect\_starvation(flow\_stat)}}
    \State \textbf{Step 8:} Mitigate starvation
    \State TCP-LLM.adjust\_CCA\_starvation(flow\_stat)
\EndIf

\State \textbf{Step 9:} Ensure CCA compatibility
\If {\texttt{detect\_incompatibility()}}
    \State \textbf{Step 10:} Select compatible CCAs
    \State TCP-LLM.select\_CCA(flow\_stat)
\EndIf

\State \textbf{Step 11:} Deploy selected CCAs
\State \texttt{deploy\_plan} = TCP-LLM.create\_plan(flow\_stat)

\State \textbf{Step 12:} Review and improve
\State \texttt{perf\_data} = \texttt{review\_perf()}
\State TCP-LLM.update\_model(perf\_data)

\end{algorithmic}
\end{algorithm}

\subsubsection{TCP-LLM Operation Algorithm}

Algorithm~\ref{design} outlines the TCP-LLM framework, a data-driven approach for managing and optimizing congestion control in TCP. The algorithm integrates convolutional Neural Networks with LLMs to dynamically predict and adjust CCA. It operates in two phases: (i) learning from historical data and (ii) real-time decision-making to ensure network fairness, prevent starvation, and achieve CCA compatibility. 

A brief explanation of Algorithm~\ref{design} steps is provided below.\\
\textbf{1. Initialization of TCP-LLM:} The framework initializes by training a model on key TCP metrics, such as throughput, loss rate, RTT, and sending rate, designed to select CCAs based on network conditions adaptively.\\
\textbf{2. Collect Historical Network Metrics:} Historical time-series data, including RTT, throughput, packet loss, and sending rates for various CCAs (e.g., Cubic, BBR, PCC), is gathered to train and refine the model.\\
\textbf{3. Encode Time Series Data Using CNN:} CNNs process time series data to extract temporal patterns, producing encoded features representing network behavior over time.\\
\textbf{4. Linear Transformation and Normalization:} Encoded features undergo linear transformation to token embeddings and normalization to ensure consistency, allowing for effective processing of network states.\\
\textbf{5. Analyze Flow Fairness:} Normalized embeddings are analyzed to identify fairness issues, ensuring bandwidth is equitably distributed across flows without disproportionate throttling.\\
\textbf{6. Minimize Flow Unfairness:} If unfairness is detected, the model dynamically adjusts CCAs to ensure equitable bandwidth allocation, addressing potential imbalances in resource distribution.\\
\textbf{7. Check and Prevent Flow Starvation:} The model detects flow starvation, where specific flows are deprived of resources, and adjusts CCA settings to restore a minimum throughput for affected flows.\\
\textbf{8. Ensure CCA Compatibility:} TCP-LLM evaluates and resolves incompatibilities between CCAs (e.g., BBR and Cubic) to prevent performance conflicts in shared environments.\\
\textbf{9. Monitor and Select Compatible CCAs:} The framework monitors congestion conditions and selects CCAs that harmonize with the network environment, ensuring stability and optimal performance.\\
\textbf{10. Deploy Selected CCAs:} Based on the analysis, the framework formulates and implements a deployment plan for transitioning to the selected CCAs without disrupting traffic.\\
\textbf{11. Feedback for Model Improvement:} After deployment, the model reviews performance data to refine future predictions and adaptations, creating a feedback loop for continuous improvement.

This algorithm offers a robust framework for dynamic and adaptive congestion control in TCP, leveraging advanced machine learning techniques to optimize performance, maintain fairness, and ensure stability across diverse network conditions.

\subsection{Integrated Encoder} 

The key to processing task-specific information such as throughput in TCP data with an LLM is to pass the TCP data input into token space, enabling proficient use by the LLM. To acquire this, we build an integrated TCP data encoder to automatically learn such projection. We have also presented an algorithm to simplify how the integrated state encoder is designed and implemented in this paper.
\begin{figure} [!h]
    \centering
    \includegraphics[scale=0.4407]{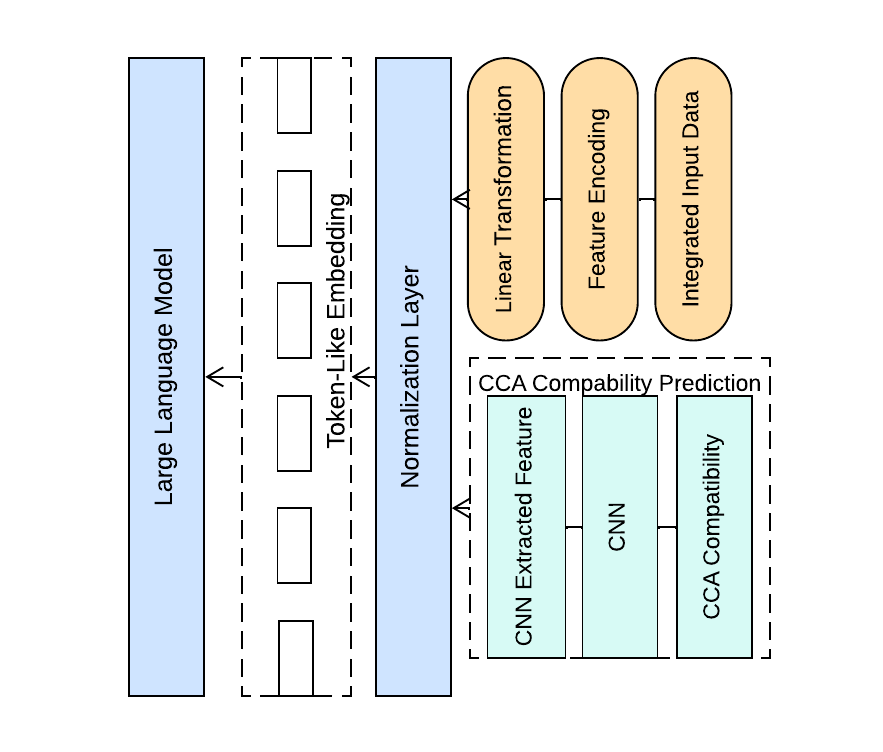}
    \caption{Graphical presentation of the integrated encoder of TCP-LLM  for TCP data encoding.}
    \label{fig:encoder}
\end{figure}

The encoder module for TCP tasks, as illustrated in Figure~\ref{fig:encoder}, is a high-efficiency architecture designed to ensure compatibility across diverse CCAs. The design integrates a 1D-CNN~\cite{bai2018empirical} to process TCP time-series data, capturing temporal dependencies effectively. Concurrently, scalar metrics like throughput and latency are encoded into embeddings through \textbf{linear projection} layers. All embeddings are layer-normalized to maintain stability and consistency, ensuring seamless integration with the LLM for downstream tasks.

This approach prioritizes modularity and proven techniques over custom designs. The 1D-CNN leverages its established strengths in temporal \textbf{feature extraction}, minimizing engineering complexity while maximizing accuracy. To bridge the gap between CNN-extracted features and the LLM’s token space, trainable linear layers map these features into compatible embeddings. \textbf{Layer normalization}~\cite{lei2016layer}  further enhances stability, enabling robust training and uniform data representation. This streamlined integration ensures a scalable and adaptable encoder tailored to the unique demands of TCP modeling.

The Algorithm~\ref{encoder} further simplifies and provides insight into the integrated encoder. 

\begin{algorithm}
\caption{TCP-LLM Integrated Encoder}\label{encoder}
\begin{algorithmic}[1]
\State \textbf{Input:} Tensor $(batch\_size, seq\_len, 4)$ of TCP metrics (Throughput, Loss Rate, RTT, Sending Rate)
\State \textbf{Output:} Encoded features $(b\_size, seq\_len, emb\_dim)$

\State \textbf{Step 1:} Extract metrics from input tensor:
\State \quad \texttt{thrup\_feat} = \texttt{state[..., 0]}
\State \quad \texttt{loss\_feat} = \texttt{state[..., 1]}
\State \quad \texttt{rtt\_feat} = \texttt{state[..., 2]}
\State \quad \texttt{send\_feat} = \texttt{state[..., 3]}

\State \textbf{Step 2:} Encode each metric via FC layers:
\State \quad \texttt{thrup\_enc} = \texttt{fc\_thrup(thrup\_feat)}
\State \quad \texttt{loss\_enc} = \texttt{fc\_loss(loss\_feat)}
\State \quad \texttt{rtt\_enc} = \texttt{fc\_rtt(rtt\_feat)}
\State \quad \texttt{send\_enc} = \texttt{fc\_send(send\_feat)}

\State \textbf{Step 3:} Reshape to original sequence format:
\State \quad \texttt{thrup\_res} = \texttt{res(thrup\_enc, (b\_size, seq\_len, emb\_dim))}
\State \quad \texttt{loss\_res} = \texttt{res(loss\_enc, (b\_size, seq\_len, emb\_dim))}
\State \quad \texttt{rtt\_res} = \texttt{res(rtt\_enc, (b\_size, seq\_len, emb\_dim))}
\State \quad \texttt{send\_res} = \texttt{res(send\_enc, (b\_size, seq\_len, emb\_dim))}

\State \textbf{Step 4:} Aggregate encoded features:
\State \quad \texttt{enc\_feat} = \texttt{concat(thrup\_res, loss\_res, rtt\_res, send\_res)}

\State \textbf{Return:} \texttt{enc\_feat}
\end{algorithmic}
\end{algorithm}

The TCP-LLM integrated encoder is a pivotal component of the TCP-LLM framework, engineered to systematically analyze and encode a sequence of critical TCP metrics over time, including Throughput, Loss Rate, RTT, and Sending Rate. Using a 3D tensor as input, structured to reflect the batch size, sequence length, and four essential TCP metrics, the algorithm efficiently extracts each metric and transforms it into a high-dimensional feature representation through a series of fully connected layers. This encoding process preserves the temporal dynamics of the metrics, allowing for a comprehensive and distinct representation of the TCP state. The encoded features are subsequently integrated to inform decision-making processes aimed at maintaining flow fairness, preventing starvation, and ensuring compatibility among various CCAs. Through this detailed encoding approach, the integrated encoder facilitates the adaptive selection of the most effective CCA through analysis of both historical and current data by enhancing the overall performance and fairness of TCP flows.

\subsection{TCP-LLM Head}

\begin{figure} [!h]
    \centering
    \includegraphics[scale=0.4407]{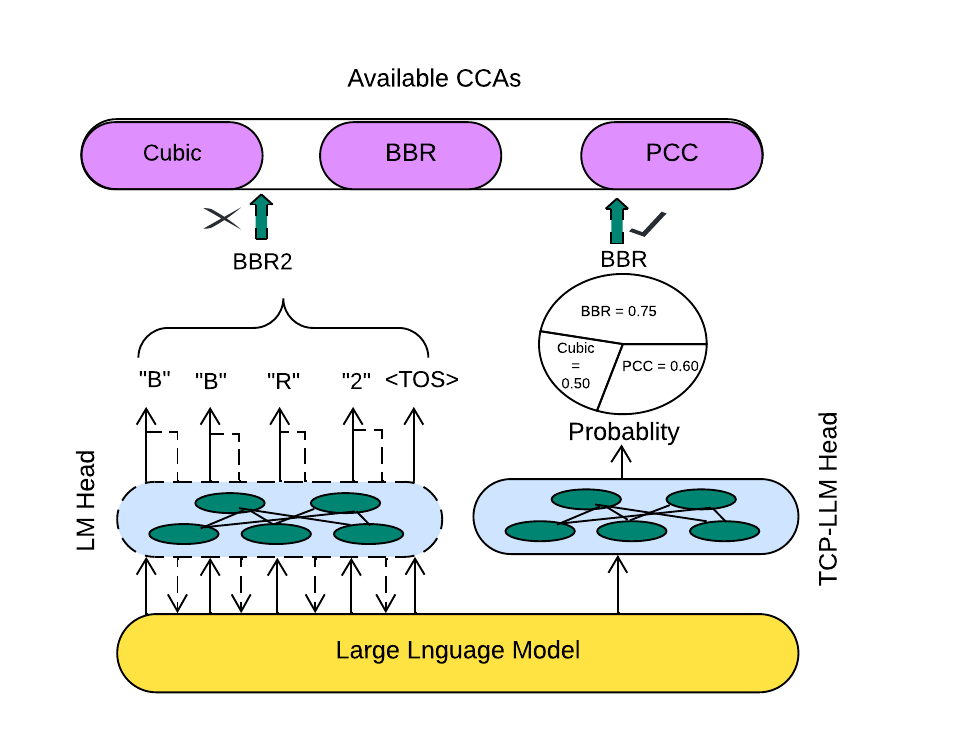}
    \caption{The detailed design of TCP-LLM Head and implementation in solving TCP-specific problems.}
    \label{fig:tcp_head}
\end{figure}

The TCP-LLM framework employs a specialized TCP data encoder to extract high-level features from key metrics like latency and congestion conditions. These features are processed by the TCP-LLM head, a trainable linear layer designed to deliver precise and context-relevant predictions for TCP tasks in a single inference round.

Unlike traditional token-based models that rely on autoregressive prediction, often requiring multiple inference rounds and risking invalid outputs like unrealistic TCP parameters, the TCP-LLM head ensures rapid and valid predictions. By focusing on a predefined set of candidate TCP parameters, it eliminates uncertainty and guarantees feasible outcomes.

For example, Figure~\ref{fig:tcp_head} highlights the efficiency of the TCP-LLM head in selecting CCAs. While traditional LM heads rely on sequential token predictions prone to delays and invalid CCAs, the TCP-LLM head directly predicts a probability distribution over valid options. This approach minimizes latency, ensures output validity, and enhances performance in TCP-specific tasks.

\begin{algorithm}[!t]
\caption{Algorithm for TCP-LLM Head}\label{head}
\begin{algorithmic}[1]
\State \textbf{Input:} 
\begin{itemize}
    \item \( input\_logits \): Tensor \( (b\_size, seq\_len, f\_dim) \) containing TCP metrics.
    \item \( input\_dim \): Features count (e.g., RTT, Throughput).
    \item \( num\_CCAs \): Available CCAs ( Cubic, BBR, PCC).
\end{itemize}

\State \textbf{Step 1:} Initialize TCP-LLM Head with parameters \( (input\_dim, num\_CCAs) \).

\State \textbf{Step 2:} Perform Forward Pass
\State \quad \( predictions = model.forward(input\_logits) \)

\State \textbf{Step 3:} (Optional) Apply Teacher Forcing
\If{teacher forcing is enabled}
    \State \( predictions = model.teacher\_forcing(input\_logits)\)
\EndIf

\State \textbf{Step 4:} Select Best CCA
\State \quad \( selected\_CCAs\_index = \texttt{argmax}(predictions) \)
\State \quad \( selected\_CCAs = ccas[selected\_CCAs\_index] \)

\State \textbf{Step 5:} Output Selected CCA
\State \Return \( selected\_CCA \)
\end{algorithmic}
\end{algorithm}

The TCP-LLM Head Algorithm~\ref{head} is designed to predict the optimal CCA based on TCP metrics. It begins by initializing the model with relevant input parameters, including the number of input features and the available CCAs. The algorithm performs a forward pass to generate predictions from the input logits. If teacher forcing is enabled, it enhances predictions for future scenarios. The algorithm then identifies the CCA with the highest predicted score by selecting the index corresponding to the maximum value in the predictions. Finally, it maps this index to the appropriate CCA and outputs the selected algorithm, ensuring efficient and accurate decision-making in TCP contexts.

\subsection{Low-Rank TCP Adaptation}

This section explains a proposed design of low-rank adaptation in TCP aimed at addressing key issues in TCP congestion control, namely flow unfairness, TCP starvation, and CCA incompatibility. The TCP adaptation process includes two central components: i) a data-driven adaptation pipeline to address TCP congestion issues, and ii) a low-rank adaptation technique to make the adaptation process more efficient.

\subsubsection{Data-Driven TCP Adaptation}

\begin{figure} [!h]
    \centering
     \includegraphics[width= 2.888 in]{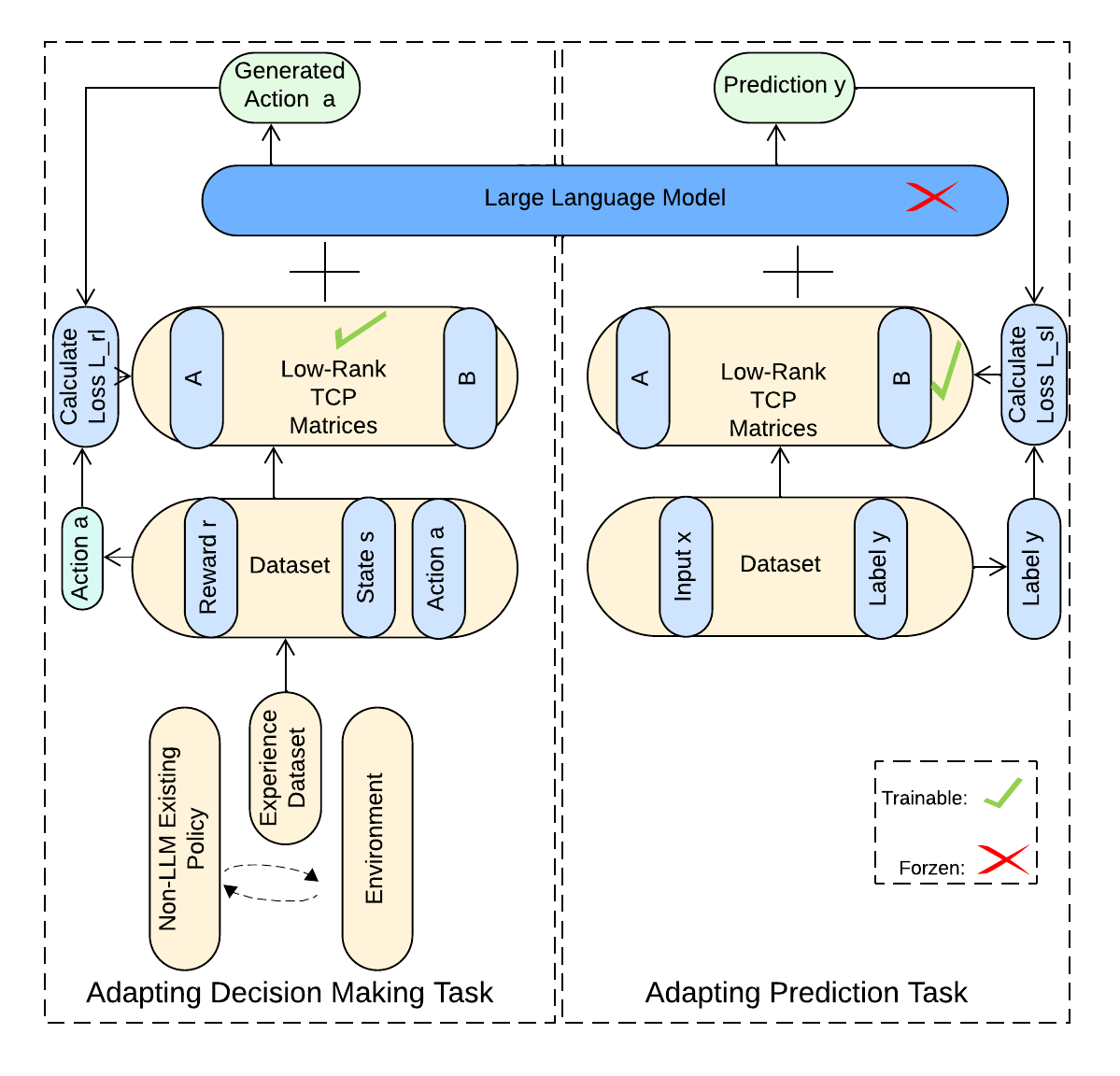}
    \caption{Visualization of Low-rank TCP Adaptation in TCP-LLM. The figure illustrates a parameter-efficient adaptation mechanism applied to both decision-making and prediction tasks. It highlights two distinct tasks: decision-making and prediction, each employing low-rank matrices to reduce computational costs while maintaining a high quality of LLMs.}
    \label{fig: approximate}
\end{figure}

\begin{algorithm}
\caption{Low-Rank Adaptation (LoRA) for TCP-LLM Fine-Tuning}\label{lora}
\begin{algorithmic}[1]
\State \textbf{Input:} Pre-trained weight matrix \( W_0 \) from a TCP-related model.
\State \textbf{Output:} Updated weight matrix \( W \) using low-rank adaptation for TCP tasks.

\State \textbf{Step 1:} Initialize low-rank matrices
\State \quad Set dimensions \( d \) (number of TCP metrics, e.g., 4), \( r \) (rank), and \( k \) (embedding dimension).
\State \quad Initialize \( A \in {R}^{d \times r} \) and \( B \in {R}^{r \times k} \) with suitable initialization (e.g., random).

\State \textbf{Step 2:} Compute the adapted weight matrix
\State \quad \( W = W_0 + AB \)

\State \textbf{Step 3:} Freeze the original parameters of the model
\For{each parameter \( p \) in the model}
    \If{\( p \) is in \( W_0 \)}
        \State Freeze \( p \)
    \EndIf
\EndFor

\State \textbf{Step 4:} Fine-tune the low-rank matrices \( A \) and \( B \)
\State \quad \texttt{for each training iteration:}
\State \quad \quad Compute the loss based on TCP tasks (e.g., throughput prediction).
\State \quad \quad Back-propagate gradients to update \( A \) and \( B \) using an optimizer (e.g., Adam).

\State \textbf{Step 5:} Return the updated model with low-rank adaptations
\State \Return{Model with \( W = W_0 + AB \) for dynamic CCA selection and performance optimization.}

\end{algorithmic}
\end{algorithm}

The Low-Rank Adaptation (LoRA) algorithm for fine-tuning the TCP-LLM model enhances the original weight matrix \( W_0 \) by integrating low-rank matrices \( A \) and \( B \). It begins by initializing these matrices with specified dimensions and ranks, then computes the adapted weight matrix \( W \) as:

\[
W = W_0 + AB.
\]

The algorithm proceeds to freeze the original model parameters to preserve the pre-trained knowledge. During training iterations, it focuses on fine-tuning the parameter efficient matrices \( A \) and \( B \) by calculating the loss related to TCP tasks and updating the matrices through backpropagation. Finally, the updated model, incorporating low-rank adaptations, is returned, facilitating dynamic CCA selection and optimizing performance in TCP-related applications.

\paragraph{Prediction Tasks:} For TCP-related prediction tasks, we utilize a standard SL data-driven training pipeline as shown in Figure~\ref{fig: approximate}. Given a dataset 
\[
D_{sl} = \{X, Y\}
\]
comprising TCP metrics such as RTT, Loss Rate, and Throughput where \( x \in X \) represents input features and \( y \in Y \) denotes labels, we leverage this dataset to fine-tune our model.

The training process involves encoding the input data \( x \) using an integrated encoder, generating prediction results \( \hat{y} \) from the model, and computing the loss function for parameter updates:
\[
L_{sl} = F_{sl}(y, \hat{y})
\]
where \( F_{sl} \) could be MSE for regression tasks, such as throughput prediction, or cross-entropy (CE) for classification tasks~\cite{pacheco2018towards}, like classifying congestion states. 

To facilitate this training, we maintain an Experience Pool that stores trajectories of states, actions, and rewards, allowing us to learn from previously encountered scenarios. Each state in the pool captures the metrics Throughput, Loss Rate, RTT, and Sending Rate that are essential for modeling TCP performance. 

The actions taken in response to these states correspond to the selected CCAs, encoded as integers. The reward function is defined to encourage efficient TCP behavior by maximizing throughput while minimizing latency and packet loss, expressed as:
\[
\text{reward} = \frac{\text{Throughput}}{\text{Latency} + 1} - \text{LossRate}
\]

\paragraph{Decision-Making Tasks} For decision-making tasks, such as dynamically adjusting TCP parameters or switching between CCAs to address issues like unfairness and starvation, the traditional RL approach can be inefficient due to extensive interaction requirements with TCP environment. To overcome this, we employ an efficient data-driven RL pipeline~\cite{liu2023constrained}.

We first collect an experience dataset using existing CCAs such as Cubic, BBR, and PCC and a policy-based system. This dataset
\[
D_{rl} = \{\tau_1, \ldots, \tau_{|D_{rl}|}\}
\]
comprises experience trajectories where each trajectory
\[
\tau = \{r_t, s_t, a_t\}_{t=1}^T
\]
includes rewards \( r \), states \( s \), and actions \( a \). We then replace the reward \( r_t \) with the return
\[
R_t = \sum_{i=t}^T r_i
\]
representing the cumulative rewards expected from state \( s_t \).

States and actions, which consist of multiple metrics (e.g., RTT, loss rate), are discretized into sequences:
\[
\tau = \{R_t, s_t, \ldots, s_t, a_t, \ldots, a_t\}_{t=1}^T 
\]

We fine-tune the model using this trajectory representation by sampling batches of inputs from the dataset, feeding to the model, computing the loss function to decide for the tasks, and updating the model parameters implementing gradient descent:
\[
L_{rl} = F_{rl}(a, \hat{a})
\]
where \( F_{rl} \) measures the difference between the actual and predicted actions, utilizing MSE for continuous actions.

\subsubsection{Low-Rank TCP Adaptation}

Given the extended parameter space of TCP models, Fine-tuning all parameters directly can be computationally prohibitive. To address this, we introduce a Low-rank TCP adaptation approach. We freeze the pre-trained parameters of the model and use Low-rank matrices~\cite{hu2021lora} to approximate parameter updates.

For each pre-trained matrix \( W_0 \), we approximate updates \( \Delta W \) using two low-rank TCP matrices \( A \) and \( B \) (see Figure~\ref{fig: approximate}):
\[
W = W_0 + \Delta W = W_0 + AB
\]
where \( A \) and \( B \) have dimensions of \( d \times r \) and \( r \times k \), respectively. Here,
\( d \) represents the number of rows in matrix \( A \),
\( r \) represents the common dimension between \( A \) and \( B \), and
\( k \) represents the number of columns in matrix \( B \).

This method drastically minimizes the trainable parameters, resulting in lower GPU memory (see Figure~\ref{fig:cost minimization}) consumption. Additionally, it shortens training time without altering the core parameters of the original model.

The low-rank approach also ensures that the model's pre-trained expertise is protected, allowing the same model to be adapted for various TCP tasks by training different Low-rank matrices \( A \) and \( B \) for each specific task. By integrating the experience pool for collecting states, actions, and rewards, we enhance the capability of the model to effectively learn and adapt, enhancing its effectiveness in real-time decision-making scenarios.

\subsubsection{In Brief}

Figure~\ref{fig: approximate} depicts a framework for efficiently adapting a pre-trained LLM to various tasks, including prediction and decision-making. The model can be fine-tuned by introducing task-specific approximate matrices and freezing the LLM's parameters to address domain-specific challenges such as fairness, starvation, and CCA compatibility. This approach allows for efficient adaptation without compromising the LLM's foundational knowledge. During training, the model processes data, calculates relevant loss functions, and updates the approximate matrices to optimize performance for the target task. This framework demonstrates the LLM's versatility and adaptability in diverse applications, from natural language processing to reinforcement learning.

\section{TCP-LLM with TCP Issues}

To address flow fairness, starvation, and CCA compatibility, TCP-LLM leverages the adaptability of LLM (Llama2-7B) in an offline setting. This approach enables dynamic analysis of congestion conditions and intelligent network adjustments, ensuring optimized and equitable performance through proactive monitoring and adaptation.

 \begin{enumerate}
     \item Flow Fairness: The TCP-LLM module monitors and manages flow fairness by dynamically selecting and deploying the most suitable CCAs based on real-time network conditions. The model ensures that network resources are distributed equitably between different flows, minimizing the likelihood that one flow dominates others.
  
     \item Prevention of starvation: TCP-LLM actively monitors the flow status and takes actions to prevent any flow from being deprived of resources. By continuously evaluating active flows' performance and adjusting the CCAs as needed, TCP-LLM ensures that each flow receives a fair share of network bandwidth, thereby preventing starvation.
    
     \item CCA Compatibility: TCP-LLM tackles CCA incompatibility by continuously monitoring the status of available CCAs (such as Reno, Cubic, PCC, and BBR) and selecting the ones that best fit the current network environment. This proactive approach helps avoid potential conflicts between different CCAs and ensures that the selected CCAs work harmoniously to optimize overall network performance.
 \end{enumerate}

 \section{Research Implementation}~\label{sec:approach}

This section describes the experimental setup and the testing methodology for evaluating TCP-LLM's performance in various network conditions. The testbed consists of client and server machines running on Ubuntu, with tools such as Iperf3, Wireshark, and Scapy used for traffic generation and data collection. Key metrics such as throughput, packet loss, and RTT are analyzed to assess the effectiveness of CCAs like CUBIC, BBR, and PCC. The training and validation of TCP-LLM and DRL model are detailed, where loss and accuracy metrics demonstrate the model's ability to make informed decisions on optimal CCA selection, improving network fairness and efficiency. The section also compares TCP-LLM's performance with state-of-the-art CCAs and 
 DRL model, showing TCP-LLM's enhanced fairness and CCA compatibility in different network scenarios.

\subsection{Experimental Setup } \label{experimental}

\begin{figure} [h]
    \centering
    \includegraphics[scale=0.19250757]{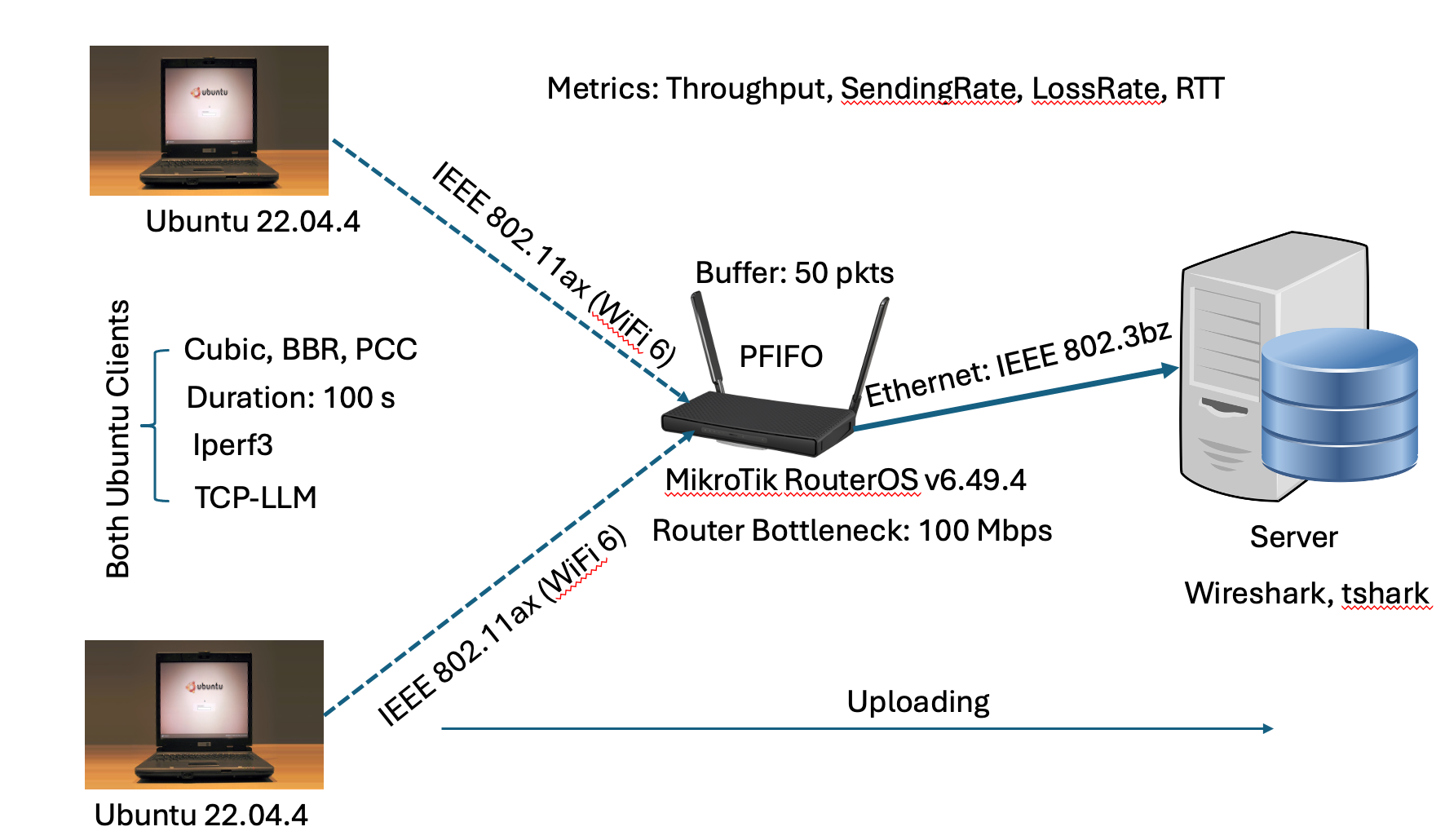}
    \caption{A block diagram of the actual lab testbed setup.}
    \label{fig:testbed}
\end{figure}

Our experimental testbed is designed to rigorously evaluate CCAs' performance under controlled network conditions. The setup includes client and server machines running Ubuntu 22.04.4, with Iperf3 employed for traffic generation and combining Wireshark, t-shark, and Scapy for comprehensive data collection. As displayed in Figure~\ref{fig:testbed}, the network backbone is established using a MikroTik RouterOS v6.49.4 router, with clients connecting via WiFi to a server through a stable Ethernet link. The network configuration introduces a 100 Mbps bottleneck and a 50-packet queue buffer managed by the Pfifo (First-In-First-Out) queuing discipline, ensuring that the algorithms are tested under bandwidth constraints and buffer management challenges.

The experiment spans 100 seconds and evaluates the performance of CCAs including CUBIC, BBR, and PCC. Key metrics such as throughput, sending rate, packet loss ratio, and RTT are analyzed to assess the algorithms' efficiency and responsiveness under varying network conditions. Additionally, the use of the WiFi hAP ax3 router introduces realistic variability, allowing a thorough investigation of the effectiveness of the CCAs in dynamic wireless environments. During the training and testing of the TCP-LLM, we used Runpod, an online platform that allows us to rent GPUs and storage for resource-intensive tasks such as model training and validation. We rented two NVIDIA A40 GPUs from Runpod to handle the heavy computational requirements.

\subsection{TCP-LLM and DRL Training and Validation}
During our experimentation with TCP-LLM, we collected various training and validation metrics, including loss, accuracy, reward, GPU utilization, computation time, and energy consumption, among others. However, to demonstrate the model’s effectiveness and utility, we have specifically presented the loss and accuracy for both the training and test phases. To provide insights into TCP-LLM's design and utility, it is essential to explain how loss and accuracy are calculated.

The loss in TCP-LLM is computed by processing each batch in the dataset, where a loss value is derived based on the model’s predictions compared to the true labels (true values). This loss is adjusted for gradient accumulation by dividing it by the specified number of accumulation steps. The adjusted loss is then used to compute and accumulate gradients through back-propagation, with gradients clipped for stability. After a defined number of steps, the model parameters are updated based on the accumulated gradients, and the gradients are reset for the next iteration. Periodically, the mean of the recorded loss values is calculated to summarize the model’s performance throughout the training process.

Similarly, accuracy is computed by comparing the model’s predicted actions, obtained via an argmax operation on \texttt{actions\_preds}, with the true actions from \texttt{actions}. The true actions are converted to class labels using a mapping function. Accuracy for each training step is determined as the fraction of correct predictions.
This results in a value between 0 and 1. The mean accuracy across all steps is then calculated to reflect the model’s overall performance for that epoch.

In comparison, our previous work involving a DRL model~\cite{shrestha2024fairness} implemented in a similar context followed a comparable approach for loss and accuracy calculations. The DRL model also relied on batch-wise loss computation with gradient accumulation and clipping for stability. Accuracy was derived by evaluating the alignment between the model's predicted actions and the true actions, similar to TCP-LLM. By benchmarking against the DRL model, we aim to showcase TCP-LLM’s advancements in efficiency and robustness for addressing the given task.

\begin{figure}[h]
    \centering
    \begin{minipage}{0.23\textwidth} 
        \centering
        \includegraphics[width=\textwidth]{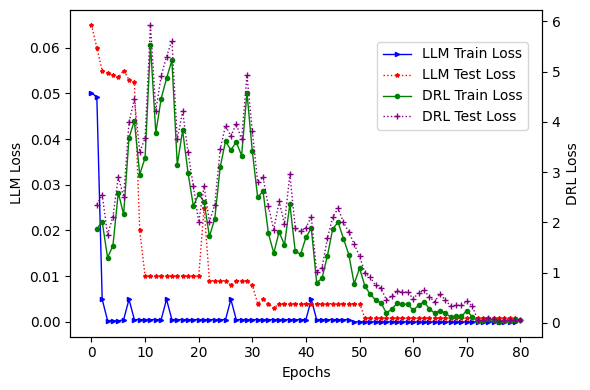}
        \caption{The evolution of training and testing loss over 80 epochs for two models TCP-LLM and DRL applied to TCP optimization tasks. The left y-axis represents the loss for TCP-LLM, while the right y-axis represents DRL loss. TCP-LLM shows stable learning with minimal fluctuations, while DRL experiences high initial loss and fluctuations, highlighting challenges in dynamic networking.}
        \label{fig:loss}  
    \end{minipage}%
    \hfill
    \begin{minipage}{0.23\textwidth} 
        \centering
        \includegraphics[width=\textwidth]{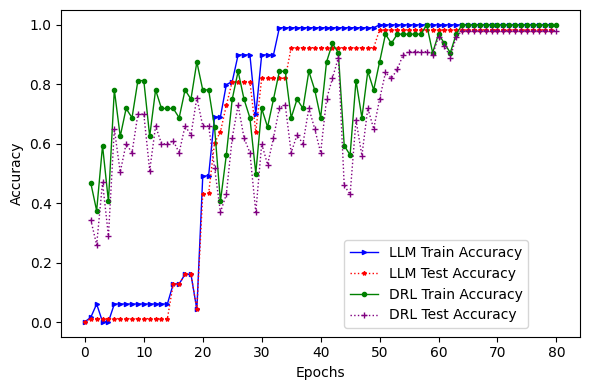}
        \caption{The illustration of the training and testing accuracy of TCP-LLM and DRL over 80 epochs for TCP optimization tasks. TCP-LLM learns quickly and achieves stable, near-perfect accuracy (1.0) by epoch 40, demonstrating dynamic adaptability. However, DRL shows slower progress and notable fluctuations, indicating difficulties in handling complex networking scenarios.}
        \label{fig:accuracy}
    \end{minipage}
\end{figure}

\subsubsection{Loss}

The Figure~\ref{fig:loss} shows the training and testing loss trajectories of TCP-LLM denoted by the blue triangle to the right (\textgreater) in a solid line and the red star (\textasteriskcentered) in a dotted line, and DRL models represented by the green circle (O) in a solid line and the purple plus (+) in a dotted line over 80 epochs, highlighting significant differences in their learning behavior and stability. The left y-axis represents the loss for TCP-LLM, while the right y-axis corresponds to the DRL loss, which operates on a significantly larger scale. TCP-LLM starts with a very low loss of about 0.06 and quickly stabilizes within a narrow range (0 to 0.06). Its learning process is efficient, achieving near-zero losses after around 5 epochs with minimal fluctuations. This consistent and stable performance demonstrates TCP-LLM’s ability to adapt quickly to complex tasks like ensuring fairness, preventing starvation, and managing congestion in TCP networks. By leveraging historical data and real-time metrics such as RTT and throughput, TCP-LLM generalizes well across dynamic conditions.

In contrast, the DRL model begins with a much higher loss above 2, which spikes further in the early epochs, reaching values over 5. Throughout the first half of training, its loss fluctuates widely between 0 and 6 before eventually stabilizing around epoch 65. These large oscillations reflect the prevalent instability and slower learning of reinforcement learning methods when applied to complex, dynamic tasks like TCP optimization. While the DRL model eventually achieves some level of stability, its slower convergence and higher overall loss indicate challenges in generalizing effectively across varying network conditions.

\subsubsection{Accuracy}

 Figure~\ref{fig:accuracy} illustrates the training and testing accuracy of TCP-LLM and DRL over 80 epochs, highlighting their effectiveness in dynamically selecting CCAs to ensure fairness, prevent starvation, and optimize resource allocation in TCP networks. TCP-LLM demonstrates a faster learning rate and superior generalization compared to DRL, achieving near-perfect accuracy earlier and with greater stability.

TCP-LLM's training and testing accuracies, represented by the triangle to the right (\textgreater) in a solid line and star (\textasteriskcentered) in a dotted line, rise sharply within the first 20 epochs, surpassing 90\% accuracy by epoch 20. By epoch 45, both accuracies plateau at near-perfect levels (1.0), showcasing TCP-LLM's ability to adapt effectively across diverse congestion scenarios. This rapid convergence reflects TCP-LLM's ability to leverage historical and real-time data for optimal CCA selection, ensuring fairness and preventing starvation under dynamic network conditions.

In contrast, DRL, represented by the circle (O) in a solid line (training) and plus (+) in a dotted line (testing), exhibits slower learning and greater variability. Although DRL's accuracy improves rapidly in the initial epochs, significant fluctuations and a generalization gap between training and testing accuracies persist until epoch 40. DRL eventually stabilizes after epoch 50, achieving accuracies around 90\% for training and slightly lower for testing. However, its slower stabilization and limited adaptability make it less effective than TCP-LLM in handling dynamic congestion scenarios.

\subsubsection{TCP-LLM vs DRL Model in TCP Tasks}

TCP-LLM demonstrates clear superiority over DRL in both loss Figure~\ref{fig:loss} and accuracy Figure~\ref{fig:accuracy} making it the better-suited model for dynamic congestion control in TCP networks. TCP-LLM achieves near-zero loss and near-perfect accuracy within 40 epochs, maintaining stability throughout training and testing. This efficient learning process highlights its capability to generalize effectively by leveraging historical data and real-time metrics like RTT and throughput. By contrast, DRL struggles with higher initial loss, larger oscillations, and slower stabilization, requiring over 60 epochs to approach comparable accuracy levels. These limitations, along with a persistent generalization gap, underscore DRL's challenges in handling complex congestion control scenarios. TCP-LLM's rapid convergence and consistent performance solidify its position as the more effective approach for optimizing TCP congestion control tasks, ensuring fairness, preventing starvation, and adapting to varying network conditions.

\subsection{LLM \& DRL over TCP and State-of-the-Art}

To evaluate the effectiveness of TCP-LLM, we compared its performance with state-of-the-art CCAs — Cubic, BBR, and PCC — under varying network conditions. The comparison focused on critical metrics such as throughput, RTT, and loss ratio, which are vital for ensuring optimal network performance. TCP-LLM's ability to dynamically select and deploy the most suitable CCAs was benchmarked against both these CCAs and the DRL~\cite{shrestha2024fairness}. The study highlights TCP-LLM's superior capabilities in maintaining flow fairness, preventing starvation, and ensuring CCA compatibility. By integrating the LLM framework into TCP-LLM, the system demonstrates significant improvements in adaptability and performance, effectively optimizing congestion control in dynamic and heterogeneous network environments.

Scenario 1: Cubic vs BBR to Cubic vs BBR$\to$ Cubic

\begin{figure}[htbp]
    \centering
    \subfloat[Throughput]{%
        \includegraphics[width=0.45150\textwidth]{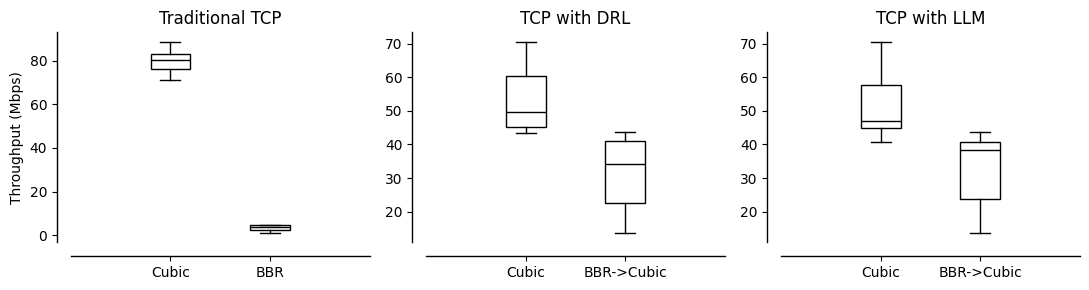}%
        \label{cubic_throughput}
    }%
    \hfill
    \subfloat[Loss]{%
        \includegraphics[width=0.451550\textwidth]{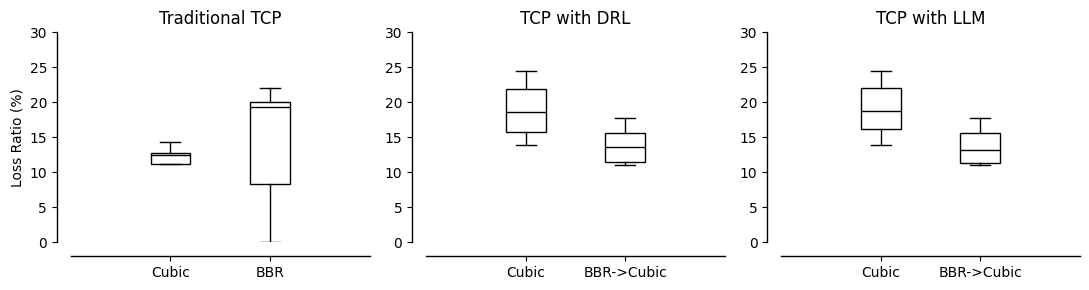}%
        \label{cubic_loss}
    }%
    \hfill
    \subfloat[RTT]{%
        \includegraphics[width=0.451550\textwidth]{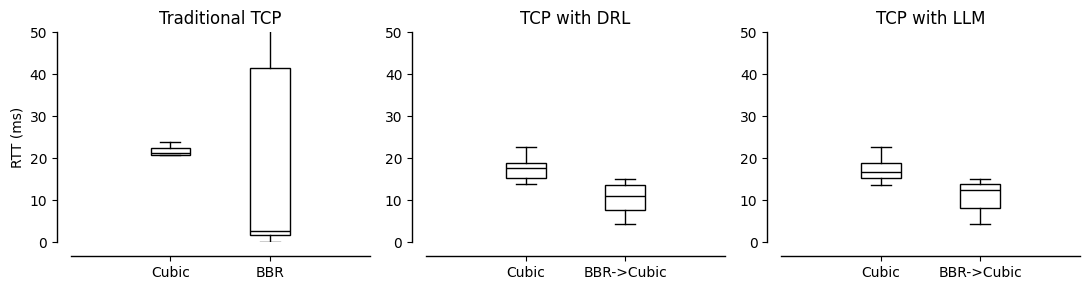}%
        \label{cubic_rtt}
    }%
    \caption{Performance metrics for Cubic vs BBR to Cubic vs BBR$\to$ Cubic                          configuration: (a) Throughput, (b) Loss, (c) RTT.}
    \label{fig:cubic_metrics}
\end{figure}

As illustrated in Figures~\ref{fig:cubic_metrics} TCP-LLM demonstrates superior performance compared to DRL and traditional CCAs like Cubic and BBR. While Cubic generally outperforms BBR in throughput (Figure~\ref{cubic_throughput}, left), loss rate (Figure~\ref{cubic_loss}, left), and RTT (Figure~\ref{cubic_rtt}, left), both exhibit significant fairness and compatibility issues under traditional TCP. Notably, Cubic achieves a median throughput of 80 Mbps compared to BBR’s 7 Mbps. DRL-based dynamic switching moderately improves these metrics, enhancing fairness and reducing inefficiencies, as shown in the middle panels of Figure~\ref{fig:cubic_metrics} for each metric. However, TCP-LLM surpasses DRL by achieving even higher throughput (42 Mbps; Figure~\ref{cubic_throughput}, right), lower loss rates (12\%; Figure~\ref{cubic_loss}, right), and reduced RTT (10 ms; Figure~\ref{cubic_rtt}, right) for BBR $\to$ Cubic transitions. These improvements underscore TCP-LLM's ability to dynamically optimize congestion control, ensuring fairness and compatibility in heterogeneous network environments.

\begin{figure}[htbp]
    \centering
    \subfloat[CDF of Throughput]{%
        \includegraphics[width=0.451880\textwidth]{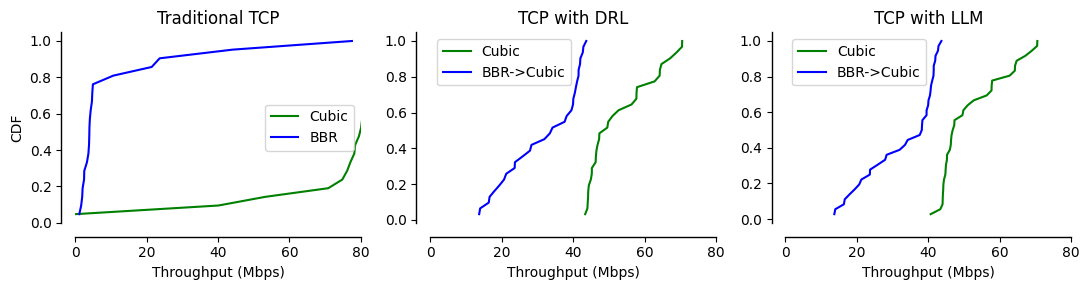}%
        \label{cdf_cubic_throughput}
    }%
    \hfill
    \subfloat[CDF of Loss]{%
        \includegraphics[width=0.45140\textwidth]{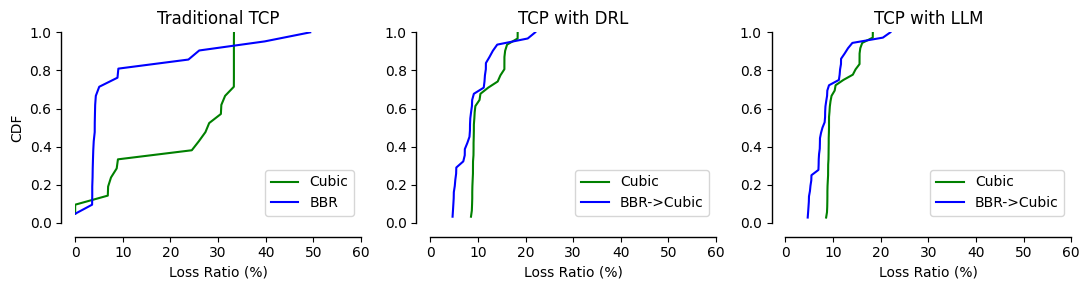}
        \label{cdf_cubic_loss}
    }%
    \hfill
    \subfloat[CDF of RTT]{%
        \includegraphics[width=0.45140\textwidth]{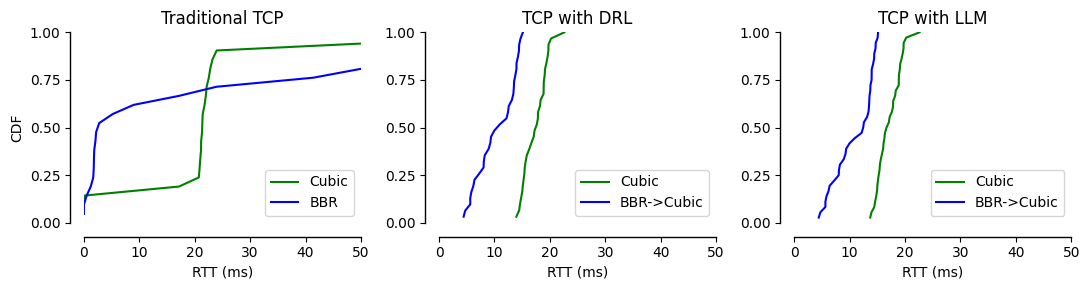}%
        \label{cdf_cubic_rtt}
    }%
    \caption{The Cumulative Distribution Function (CDF) of performance metrics, including (a) CDF of Throughput, (b) CDF of Loss, and (c) CDF of RTT, is shown for the following scenarios: state-of-the-art CCAs (Cubic vs. BBR), TCP with DRL, and TCP with TCP-LLM. The analysis specifically examines the transition from Cubic vs BBR to Cubic vs BBR$\to$ Cubic highlighting the comparative performance across methods.}
    \label{fig:cdf_cubic_metrics}
\end{figure}

The Figures~\ref{cdf_cubic_throughput}, ~\ref{cdf_cubic_loss}, and ~\ref{cdf_cubic_rtt} present CDFs for throughput, packet loss, and RTT respectively focusing on the transition from Cubic to BBR and back to Cubic (Cubic vs BBR $\to$ Cubic). We can see that the findings in Figures~\ref{fig:cdf_cubic_metrics} verify our findings discussed in Figure~\ref{fig:cubic_metrics}, offering deeper insights into the adaptability and fairness with TCP-LLM. By visualizing these metrics, the CDFs provide a holistic perspective on how TCP-LLM enhances network performance compared to Traditional TCP and TCP with DRL. The results underscore TCP-LLM's effectiveness in achieving more balanced throughput, reducing loss rates, and maintaining lower RTT across varying congestion control algorithms.

Scenario 2: PCC vs BBR to PCC vs BBR $\to$ PCC

\begin{figure}[htbp]
    \centering
    \subfloat[Throughput]{%
        \includegraphics[width=0.451550\textwidth]{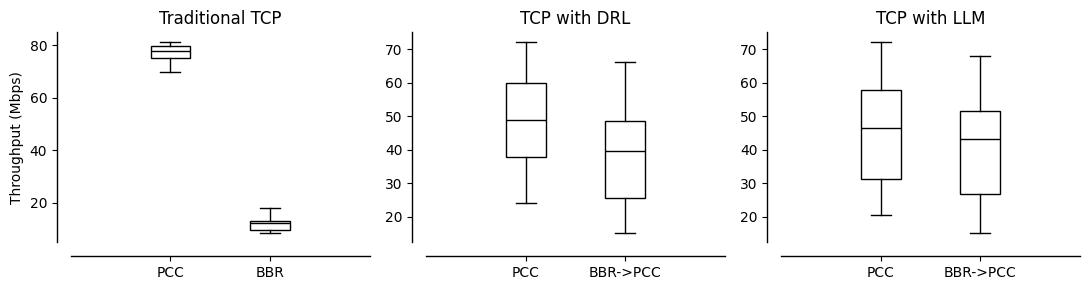}%
        \label{pcc_throughput}
    }%
    \hfill
    \subfloat[Loss]{%
        \includegraphics[width=0.451550\textwidth]{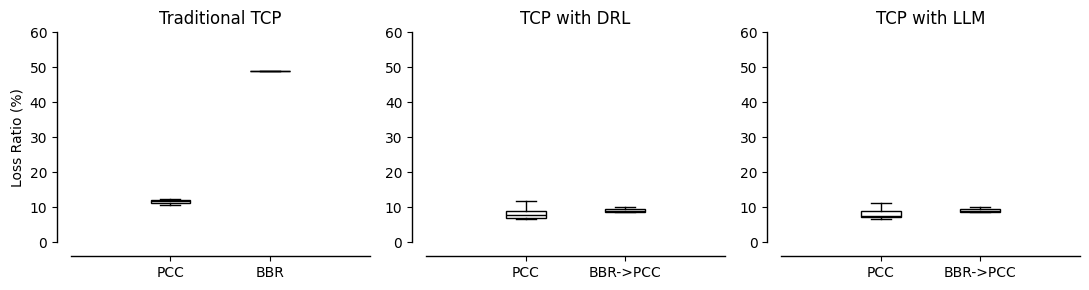}%
        \label{pcc_loss}
    }%
    \hfill
    \subfloat[RTT]{%
        \includegraphics[width=0.451550\textwidth]{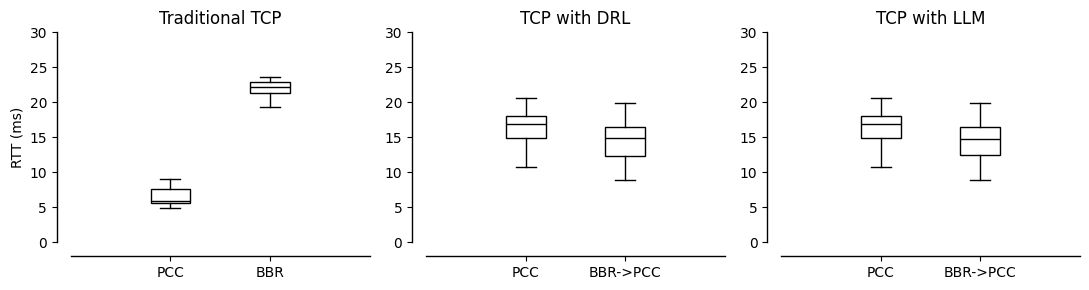}%
        \label{pcc_rtt}
    }%
    \caption{Throughput, RTT, and loss comparison of PCC vs BBR to PCC vs BBR $\to$ PCC.}
    \label{fig:pcc_metrics}
\end{figure}

Figures~\ref{pcc_throughput},~\ref{pcc_loss}, and ~\ref{pcc_rtt} compare the performance of PCC and BBR under Traditional TCP, a DRL-based model, and our TCP-LLM across metrics: throughput (Figure~\ref{pcc_throughput}), loss  (Figure~\ref{pcc_loss}), and RTT  (Figure~\ref{pcc_rtt}), demonstrating how TCP-LLM achieves better fairness, reduces incompatibility issues, and improves network performance.

Under traditional TCP, PCC significantly outperforms BBR in throughput, achieving 75 Mbps compared to BBR’s 10 Mbps (see Figure~\ref{pcc_throughput} left). This disparity highlights clear fairness and compatibility issues between the two CCAs. Additionally, BBR suffers from much higher loss rates (50\% see Figure~\ref{pcc_loss} left) and latency (25 ms see Figure~\ref{pcc_rtt} left) compared to PCC, which shows minimal losses (10\%) and lower RTT (7 ms). These findings emphasize the challenges in balancing performance across CCAs when using Traditional TCP.

The DRL-based model improves fairness and performance by dynamically transitioning from underperforming CCAs to more reliable alternatives. In the BBR→PCC configuration, the DRL model achieves a balanced throughput of 50 Mbps vs 40 Mbps (see Figure~\ref{pcc_throughput} middle). Loss rates are also reduced to 10\% (see Figure~\ref{pcc_loss} middle), while RTT stabilizes at 15 ms (see Figure~\ref{pcc_rtt} middle), showing a significant improvement over Traditional TCP. These results demonstrate how DRL addresses the compatibility gap between PCC and BBR while improving overall fairness.

The TCP-LLM model builds upon this foundation and delivers even better results. Throughput remains more balanced at 45 Mbps (see Figure~\ref{pcc_throughput} right) but shows reduced variability, ensuring more consistent performance. Loss rates are further balanced to 8\% (see Figure~\ref{pcc_loss} right), and RTT remains stable at 15 ms (see Figure~\ref{pcc_rtt} right)  with less fluctuation. This stability highlights TCP-LLM’s ability to better adapt to varying network conditions while maintaining fairness and efficiency.

\begin{figure}[htbp]
    \centering
    \subfloat[CDF of Throughput]{%
        \includegraphics[width=0.451880\textwidth]{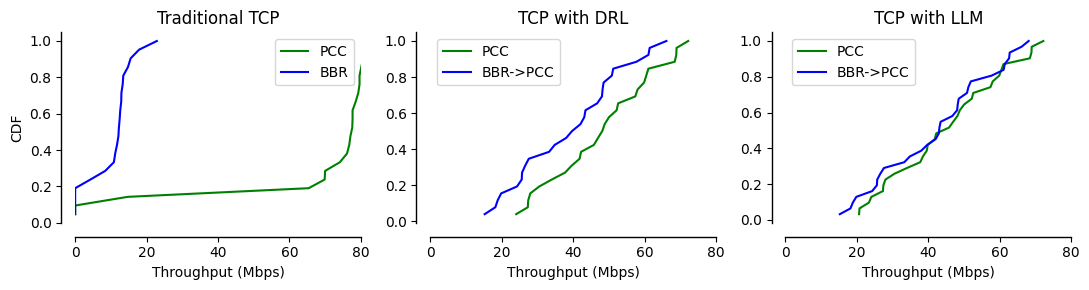}%
        \label{cdf_pcc_throughput}
    }%
    \hfill
    \subfloat[CDF of Loss]{%
        \includegraphics[width=0.45140\textwidth]{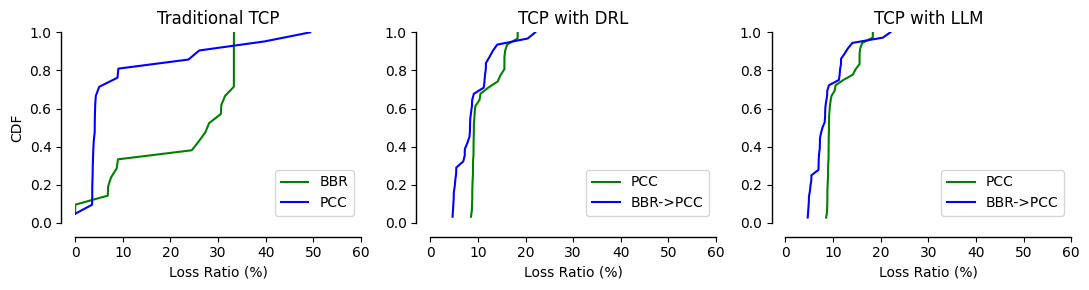}%
        \label{cdf_pcc_loss}
    }%
    \hfill
    \subfloat[CDF of RTT]{%
        \includegraphics[width=0.45140\textwidth]{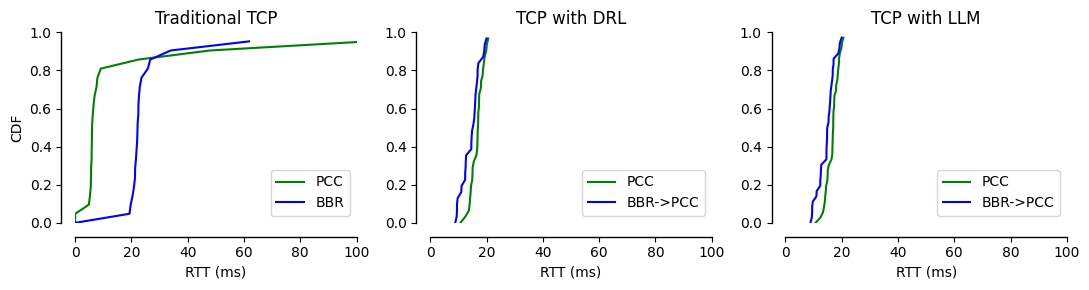}%
        \label{cdf_pcc_rtt}
    }%
    \caption{Illustration of the CDF for key performance metrics (a) CDF of Throughput, (b) CDF of Loss, and (c) CDF of RTT across three paradigms: PCC vs. BBR, TCP augmented with DRL, and TCP-LLM. The analysis emphasizes the transition dynamics from PCC vs BBR to PCC vs BBR $\to$ PCC, highlighting the comparative efficiency and adaptability of each approach.}
    \label{fig:cdf_pcc_metrics}
\end{figure}

The Figure~\ref{cdf_pcc_throughput}, \ref{cdf_pcc_loss}, and \ref{cdf_pcc_rtt} illustrates the CDFs of throughput, packet loss, and RTT simultaneously for a scenario involving transitions between PCC and BBR, specifically PCC $\to$ BBR $\to$ PCC. These CDFs provide an alternative representation of performance dynamics, capturing the impact of adaptive methods in Figure~\ref{fig:pcc_metrics}. The results reemphasize the capability of TCP-LLM to stabilize throughput, minimize loss, and ensure fairness across CCA transitions, particularly in comparison to Traditional TCP and TCP with DRL. This visualization effectively highlights the enhanced consistency and adaptability of TCP-LLM in managing dynamic CCA interactions.

Scenario 3: BBR vs PCC to BBR vs PCC$\to$ BBR

\begin{figure}[htbp]
    \centering
    \subfloat[Throughput]{%
        \includegraphics[width=0.451550\textwidth]{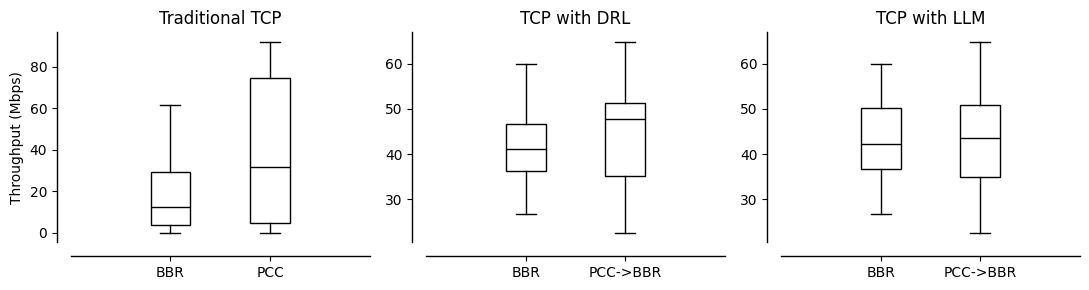}%
        \label{bbr_throughput}
    }%
    \hfill
    \subfloat[Loss]{%
        \includegraphics[width=0.451550\textwidth]{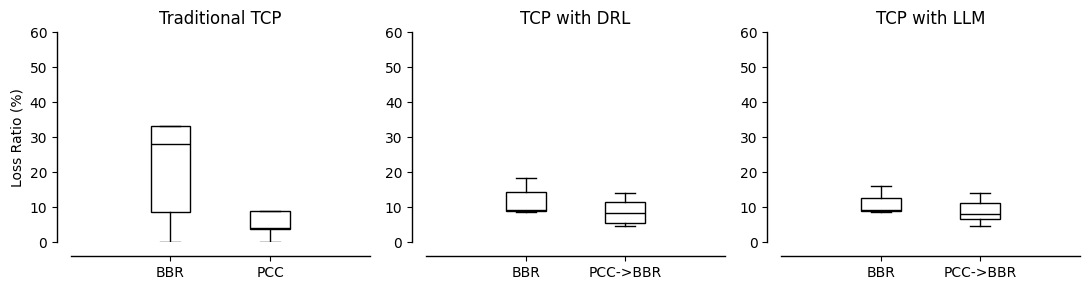}
        \label{bbr_loss}
    }%
    \hfill
    \subfloat[RTT]{%
        \includegraphics[width=0.451550\textwidth]{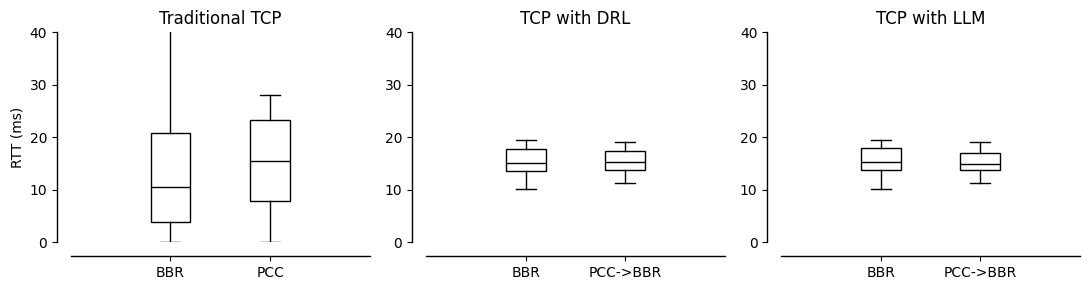}%
        \label{bbr_rtt}
    }%
    \caption{Performance metrics for BBR vs PCC to BBR vs PCC$\to$ BBR: (a) Throughput, (b) Loss, (c) RTT.}
    \label{fig:bbr_metrics}
\end{figure}

The Figure~\ref{fig:bbr_metrics}, demonstrates the performance of BBR and PCC across Traditional TCP, TCP with DRL, and TCP-LLM configurations, highlighting the progression in fairness and efficiency.

In the Traditional TCP setup, PCC achieves significantly higher throughput 40 Mbps vs 10 Mbps (see Figure~\ref{bbr_throughput} left) for BBR with lower loss rates 10\% vs 30\% (see Figure~\ref{bbr_loss} left) and RTT 10 ms vs 18 ms (see Figure~\ref{bbr_rtt} left), underscoring severe CCA incompatibility and unfairness.

The DRL model mitigates these issues, balancing throughput above 40 Mbps (see Figure~\ref{bbr_throughput} middle) for both CCAs while reducing BBR’s loss rate to ~15\% (see Figure~\ref{bbr_loss} middle) and RTT to ~18 ms (see Figure~\ref{bbr_rtt} middle), demonstrating substantial fairness improvements. Refer to \cite{shrestha2024fairness} for details.

TCP-LLM further refines performance, maintaining a similar throughput of 40 Mbps (see Figure~\ref{bbr_throughput} right) while achieving lower loss rates at 12\% (see Figure~\ref{bbr_loss} right) and more stable RTT of 15 ms(see Figure~\ref{bbr_rtt} right), demonstrating superior alignment and fairness across CCAs. This highlights TCP-LLM as a robust solution for addressing CCA compatibility challenges.

\begin{figure}[htbp]
    \centering
    \subfloat[CDF of Throughput]{%
        \includegraphics[width=0.451880\textwidth]{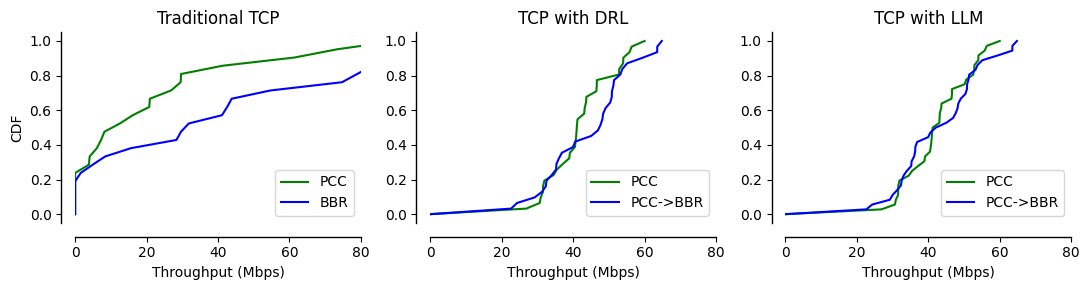}%
        \label{cdf_bbr_throughput}
    }%
    \hfill
    \subfloat[CDF of Loss]{%
        \includegraphics[width=0.45140\textwidth]{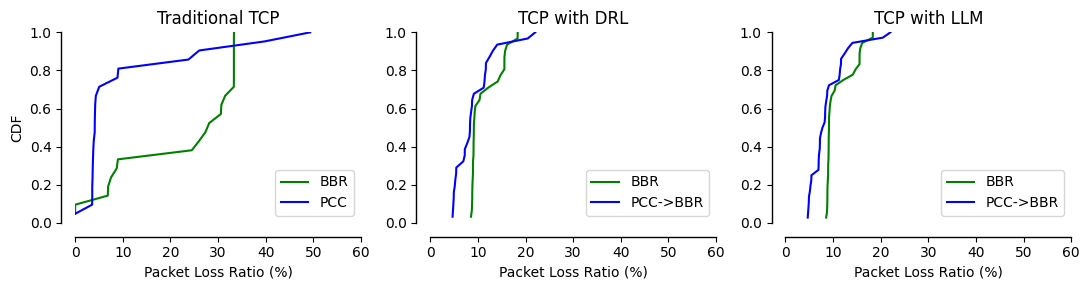}%
        \label{cdf_bbr_loss}
    }%
    \hfill
    \subfloat[CDF of RTT]{%
        \includegraphics[width=0.45140\textwidth]{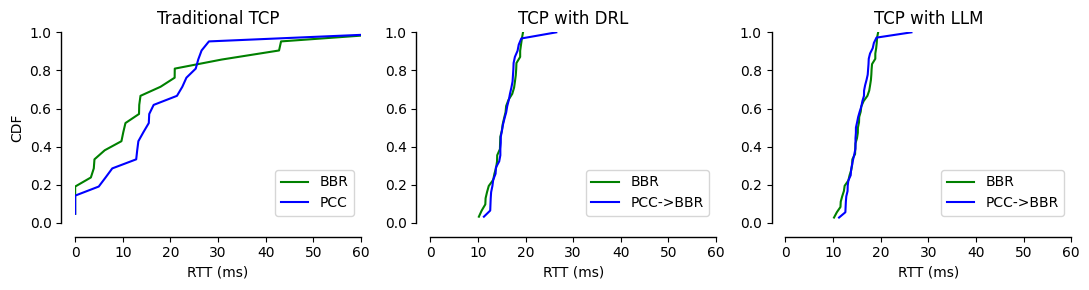}%
        \label{cdf_bbr_rtt}
    }%
    \caption{CDF of performance metrics (a) CDF of Throughput, (b) CDF of Loss, and (c) CDF of RTT comparing BBR vs. PCC, TCP with DRL, and TCP with TCP-LLM in the scenario transitioning from BBR vs PCC to BBR vs PCC$\to$ BBR.}

     \label{fig:cdf_bbr_metrics}
\end{figure}

The Figures~\ref{cdf_bbr_throughput}, \ref{cdf_bbr_loss}, and \ref{cdf_bbr_rtt} present the CDFs of throughput, packet loss, and RTT for BBR and PCC across three configurations: Traditional TCP, TCP with DRL, and TCP-LLM. These CDFs complement the previously discussed box plots~\ref{fig:bbr_metrics}, offering an alternative perspective on the same results. The trends reinforce the earlier findings, showcasing the improvements in fairness, efficiency, and stability achieved through adaptive approaches, particularly with TCP-LLM, which demonstrates superior consistency across all metrics compared to Traditional TCP and TCP with DRL. This alternative visualization highlights the robustness of the adaptive frameworks in mitigating CCA incompatibilities.

\subsubsection{Summary of State-of-the-art CCAs vs TCP with DRL vs TCP with TCP-LLM}

TCP-LLM outperforms both DRL-based models and traditional CCAs like Cubic, BBR, and PCC by dynamically optimizing congestion control, ensuring fairness, and resolving CCA compatibility issues. Compared to Traditional TCP, it achieves higher throughput, lower loss rates, and reduced RTT across all tested scenarios, including transitions between CCAs. While DRL improves fairness and mitigates inefficiencies, TCP-LLM delivers more consistent performance with reduced variability and superior adaptability under dynamic network conditions. The results, supported by detailed metric analyses and CDF visualizations, highlight TCP-LLM’s effectiveness as a robust, adaptive solution for addressing fairness, efficiency, and compatibility challenges in heterogeneous network environments.

\section{Empirical Evaluation} \label{sec:evaluation}

This section evaluates the proposed TCP-LLM framework against DRL benchmarks in addressing fairness, starvation, and CCA incompatibility in TCP tasks. We demonstrate TCP-LLM's computational efficiency, real-time adaptability, and superior generalization capabilities.

\subsection{Overcoming DRL Challenges in TCP Tasks with LLMs}

TCP-LLM outperforms DRL models in critical aspects, showcasing significant advancements in efficiency and adaptability:

\begin{itemize}
    \item \textbf{Enhanced Generalization and Adaptability:} Unlike DRL, which requires extensive retraining for each new network scenario, TCP-LLM leverages pre-trained capabilities to dynamically adapt to unseen conditions without retraining. This enables TCP-LLM to maintain robust performance across diverse environments, ensuring fairness and effectively resolving CCA compatibility issues. Figures~\ref{fig:cubic_metrics}, \ref{fig:pcc_metrics}, and \ref{fig:bbr_metrics} highlight TCP-LLM’s ability to balance throughput and reduce loss rates across heterogeneous conditions. In contrast, DRL models struggled with fluctuating RTT, unstable throughput, and higher loss rates, underscoring TCP-LLM's superior adaptability and reliability.
    
    \item \textbf{Reduced Computational Costs:} DRL's iterative training processes, marked by slow convergence and large oscillations in loss Figure~ \ref{fig:loss} and accuracy Figure~\ref{fig:accuracy}, significantly increase computational demands. Extended training times result in higher resource utilization, energy consumption, and delays, making DRL less practical for real-time applications. By integrating Low-Rank TCP Adaptation (LoRaTCPA), TCP-LLM reduces trainable parameters by 99\%, leading to a GPU memory usage of just 28.23 GB (see Figure~\ref{fig:cost minimization}). TCP-LLM's faster convergence and stable learning dynamics further minimize computational overhead, ensuring it is both resource-efficient and suitable for real-time scenarios.
    
    \item \textbf{Real-Time Decision-Making:} TCP-LLM achieves a single-step inference response time of 0.015 seconds, ensuring rapid decision-making with 100\% accuracy (see Figures~\ref{fig:answer_gen} and \ref{fig:answer_val}  Additionally, TCP-LLM demonstrates superior adaptability to dynamic network environments, resulting in consistently lower RTT values compared to DRL as seen in the Figures~\ref{cubic_rtt}, \ref{pcc_rtt}, and \ref{bbr_rtt}. The higher RTT observed in DRL implementations underscores its slower responsiveness and reduced efficiency in real-time scenarios, further highlighting TCP-LLM's advantages for tasks requiring immediate adjustments.
\end{itemize}

\subsection{LLM Adaptation for TCP Tasks}

Adapting LLMs for TCP tasks introduced unique challenges compared to DRL benchmarks, which are inherently designed for structured numerical data. TCP-LLM addresses these challenges through the following innovations:

\begin{itemize}
    \item \textbf{Modality Alignment:} TCP-LLM employs an integrated encoder (Figure~\ref{fig:encoder}) to transform TCP metrics (RTT, throughput, loss rate) into token-like embeddings. This enables the effective processing of TCP metrics akin to textual data, bridging the modality gap seamlessly.
    
    \item \textbf{Resource Optimization:} The Low-Rank TCP Adaptation Figure~\ref{fig: approximate} reduces trainable parameters by 99\%, cutting GPU memory usage from 65.88~GB (Llama2) to 28.23~GB (see Figure~\ref{fig:cost minimization}). This optimization lowers operational costs and accelerates model adaptation.
    
    \item \textbf{Real-Time Responsiveness:} TCP-LLM generates responses in a single inference step with a latency of 0.015 seconds (see Figure~\ref{fig:tcp_head}). DRL models, in contrast, exhibit higher latency due to iterative policy updates as in Figures~\ref{cubic_rtt}, \ref{pcc_rtt}, and \ref{bbr_rtt}.
\end{itemize}

\subsection{Addressing TCP Unfairness in Heterogeneous Wi-Fi Networks}

TCP-LLM significantly enhances fairness in heterogeneous Wi-Fi networks by dynamically adjusting CCAs based on real-time network conditions. For instance, Figure~\ref{fig:cubic_metrics} shows that TCP-LLM balanced Cubic and BBR flows at 45~Mbps and 40 Mbps (Figure~\ref{cubic_throughput}), reducing BBR’s packet loss from 20\% to 12\%~(Figure~\ref{cubic_loss}). DRL benchmarks, however, capped Cubic throughput at 52~Mbps while BBR achieved 37~Mbps, failing to stabilize RTT and loss metrics under similar conditions.

TCP-LLM’s rapid convergence and stable performance are further illustrated in Figures~\ref{fig:loss} and \ref{fig:accuracy}. Within 40 epochs, TCP-LLM achieves near-zero loss and near-perfect accuracy (1.0), while DRL requires over 50 epochs to stabilize with significant oscillations. These results further indicates TCP-LLM's effectiveness in ensuring fairness, preventing starvation, and addressing CCA compatibility issues.

\section{Conclusion}~\label{sec:conclusion}
In this work, we introduced the application of LLMs as foundational models for TCP tasks, aiming to minimize the manual effort in algorithm design while achieving robust generalization across varied scenarios. To realize this vision, we designed TCP-LLM, the first framework to efficiently adapt LLMs for addressing diverse TCP-specific challenges. Through multiple use cases, we demonstrate that TCP-LLM can leverage one LLM to enhance TCP performance and model generalization across TCP tasks, including congestion control, flow fairness, and CCA selection. Although TCP-LLM is not the definitive solution, we hope it catalyzes a more adaptive and scalable approach to TCP algorithm design, showcasing the immense potential of LLMs in transforming TCP performance optimization.

\bibliographystyle{plain}
\bibliography{Shyam}

 \end{document}